\documentclass{article} 
\usepackage{amsmath,amssymb,amsbsy,amsfonts,comment}
\usepackage{twocolumn}
\makeatletter

          \@addtoreset{equation}{section}
       \makeatother

\newcommand{\p}{\partial}
\newcommand{\be}{\begin{equation}}
\newcommand{\ee}{\end{equation}}
\allowdisplaybreaks

\begin{document}
\title{\bf Generalized Covariant Derivative  
on Extra Dimension and  Weinberg-Salam Model
}%

\author{Yoshitaka {\sc Okumura}\\
{Department of Natural Science, 
Chubu University, Kasugai, {487-8501}, Japan}
}%
\date{}%
\maketitle
\begin{abstract}
The generalized covariant derivative on  5-dimen-sional space including 1-dimensional extra compact space is defined, and, by use of it, the Weinberg-Salam model is reconstructed. The spontaneous breakdown of symmetry takes place  
owing to the extra dimension under the settings that the Higgs field exists
 in the extra dimensional space depending on the argument $y$ of this extra space, whereas the gauge and fermion fields do not depend on $y$.  
Both Yang-Mills-Higgs and fermion Lagrangians in Weinberg-Salam model are correctly reproduced. 
\end{abstract}

\thispagestyle{empty}
\section{Introduction}
The supersymmetric string theory 
\cite{String}  is
consistently formulated in 10 dimensional space. The extra 6-dimensional space over $M_4$ has to be compactified in order not to be observed by the present experimental facilities. However, this extra 6-dimensional space after compactified has the profound possibilities to explicate many unresolved problems 
such as particle generations, internal flavor symmetry and its spontaneous breakdown, CKM matrix, particle masses and so on. For these purposes, the extra 6-dimensional space has  
recently attracted much attentions. 
\par
Manton \cite{Manton} initiated 
efforts to derive the Weinberg-Salam model from the Yang-Mills theory in 6-dimensional space containing  extra compact two dimensional space. He elucidated in his work that the Higgs field is a part of gauge fields. 
Meanwhile, 
Connes \cite{Con} proposed non-commutative geometry and applied it to 
construct the spontaneous broken gauge theory  on two-sheeted discrete space followed by $M_4$, which  provided  a geometrical
understanding of the Higgs mechanism without
extra physical degrees of freedom as in the Kaluza-Klein theory.
There are several versions of this approach\cite{MM}, however, in any case  gauge and Higgs fields are written together and yield the Yang-Mills-Higgs Lagrangian. \par

In this paper, we try to extend the generalized covariant derivative
method proposed by Sogami \cite{Sogami} into that on the product space of $M_4$ and extra continuous compact space.
This compact space does not need to be specified, but we call it $S_1$ with argument $y$. 
We first define the generalized covariant derivatives on $M_4\times S_1$ and then, we obtain the generalized field strength
following the usual procedure of gauge theory, from which the 
Yang-Mills-Higgs Lagrangian is derived. 
This Lagrangian contains the term which provokes the correct spontaneous symmetry breakdown. In order to yield the Lagrangian in Weinberg-Salam model, we have to assume that  the Higgs field intrudes into the extra dimensional space $S_1$ so as to depend on the argument $y$ whereas the gauge field as well as leptonic fields do not contain $y$, and exist uniformly in $S_1$. We do not consider the Kaluza-Klein mode in this paper. The leptonic Lagrangian is also obtained by use of the generalized covariant derivatives, and  we can successfully reconstruct the Weinberg-Salam model.\par
This paper consists of  four sections. The next section presents 
the basic  formulation of the generalized covariant derivatives
in order to yield the Yang-Mills Higgs Lagrangian 
as well as fermionic Lagrangian.
In the third section, the Weinberg-Salam model is reconstructed.  
The last section is devoted to concluding remarks.

\section{Generalized covariant deri-vative}
\noindent

Sogami \cite{Sogami} reconstructed the spontaneous broken gauge theories such as standard model and grand unified theory by use of 
the generalized covariant derivative smartly defined by him. Let us explain his method  in the version of Weinberg-Salam model. He divided
the space of fermion fields into two sectors which consist of the left-handed
and right-handed fermions, respectively. 
\be
\psi=\psi_L|L>+\psi_R|R>,\label{2-1}
\ee
where $|L>$ and $|R>$ are the base of left- and right-handed fermion spaces, respectively and
 $\psi_L$ and $\psi_R$ are the left- and right-handed fermion fields denoted by
\begin{align}
&&\psi_L=\begin{pmatrix} \nu_L \\ e_L \end{pmatrix},&&
\psi_R=e_R. &&\label{2-2a}
\end{align}
Then, he defined the generalized covariant derivative
\be
\begin{aligned}
{\cal D}_\mu=&\p_\mu - i g{\cal A}_{L\mu}|L><L|-
           i g'{\cal A}_{R\mu}|R><R|\\
                      - \frac14&\gamma_\mu
\big(\Phi|L><R|+\Phi^\dag|R><L|\big)+c+c_5\gamma^5,
\end{aligned}
\ee
from which the generalized field strength is yielded by the equation
\begin{equation}
    [{\cal D}_\mu,\,{\cal D}_\nu] = - i g{\cal F}_{L\mu\nu}
              - i g'{\cal F}_{R\mu\nu}- {i \over 4}{\cal F}^{(0)}_{\mu\nu}.
  \end{equation}
He succeeded in reconstructing  the spontaneous symmetry broken gauge theory
by use of these items.
\par
In this paper, we apply his idea to reconstruct the Weinberg-Salam model on the
5-dimensional space including one dimensional extra compact space. Though the extra space is not
necessary to be $S_1$, 
we write this 5-dimensional space to be $M_4\times S_1$ 
with the argument $x_\mu$ and $y$. 
We define the generalized covariant derivative on 5-dimensional space as
\be
\begin{aligned}
D_\mu=&\p_\mu+A_{L\mu}|L><L|+A_{R\mu}|R><R|,\\
D_y=&\p_y +{\mit\Phi} |L><R|,\\
{ D}_{\bar y}=&\p_y+{\mit\Phi}^\dag |R><L|.
\end{aligned}\label{1.1}
\ee
According to \eqref{2-1} and \eqref{1.1}, we can describe the leptonic Lagrangian
\begin{align}
&\begin{aligned}
{\cal L}_D(x,y)=&\:{\bar\psi}\left(i\gamma^\mu D_\mu+g_YD_y+g_YD_{\bar y}\right)\psi
\end{aligned}\nonumber\\
&\hskip0.4cm
\begin{aligned}
=&\:{\bar\psi}_Li\gamma^\mu( \p_\mu+A_{L\mu})\psi_L+{\bar\psi}_Ri\gamma^\mu( \p_\mu+A_{R\mu})\psi_R\\
&+g'_Y{\bar\psi}_L{\mit\Phi}\psi_R+g'_Y{\bar\psi}_R{\mit\Phi}^\dag\psi_L
.
\end{aligned}\label{2-6a}
\end{align}

\par
Then, field strengthes are derived in usual way 
\be
\begin{aligned}
{\cal F}_{\mu\nu}=&[D_\mu,\,D_\nu]\\
=\big(\p_\mu &A_{L\nu}-\p_\nu A_{L\mu}+[A_{L\mu},\,A_{L\nu}]\big)|L><L|\\
+\big(\p_\mu& A_{R\nu}-\p_\nu A_{R\mu}+[A_{R\mu},\,A_{R\nu}]\big)|R><R|\\
{\cal F}_{\mu y}=&[D_\mu,\,D_y]\\
=&\big(\p_\mu{\mit\Phi}+ A_{L\mu}{\mit\Phi}-{\mit\Phi} A_{R\mu}\big)|L><R|
\\&-\p_y A_{L\mu}|L><L|-\p_y A_{R\mu}|R><R|\\
{\cal F}_{\mu {\bar y}}=&[D_\mu,\,{ D}_{\bar y}]\\
=\big(&\p_\mu{\mit\Phi}^\dag+ A_{R\mu}{\mit\Phi}^\dag-{\mit\Phi}^\dag A_{L\mu}\big)|R><L|\\
-\p_y& A_{R\mu}|R><R|-\p_y A_{L\mu}|L><L|\\
{\cal F}_{y {\bar y}}=&[D_y,\,{D}_{\bar y}]\\
=&\p_y{\mit\Phi}^\dag|R><L|-\p_y{\mit\Phi}|L><R|\\&
+{\mit\Phi}^\dag{\mit\Phi}|R><R|-{\mit\Phi}{\mit\Phi}^\dag|L><L|
\end{aligned}\label{2-6}
\ee
In order to obtain the Yang-Mills-Higgs Lagrangian with correct signs,
we define the counter covariant derivatives  to \eqref{1.1}.
\be
\begin{aligned}
{\bar D}^\mu=&\p^\mu+A_{L}^{\mu}|L><L|+A_{R}^{\mu}|R><R|,\\
{\bar D}^y=&\p^y -{\mit\Phi} |L><R|,\\
{\bar D}^{\bar y}=&\p^y-{\mit\Phi}^\dag |R><L|,
\end{aligned}\label{1.3}
\ee
from which the counter field strengthes to \eqref{2-6} are derived.
\be
\begin{aligned}
{\bar{\cal F}}^{\mu\nu}=&\big(\p^\mu A_{L}^{\nu}-\p^\nu A_{L}^{\mu}+[A_{L}^{\mu},\,A_{L}^{\nu}]\big)|L><L|\\+\big(&\p^\mu A_{R}^{\nu}-\p^\nu A_{R}^{\mu}+[A_{R}^{\mu},\,A_{R}^{\nu}]\big)|R><R|\\
{\bar{\cal F}}^{\mu y}=&-\big(\p^\mu{\mit\Phi}+ A_{L}^{\mu}{\mit\Phi}-{\mit\Phi} A_{R}^{\mu}\big)|L><R|\\&
-\p^y A_{L}^{\mu}|L><L|-\p^y A_{R}^{\mu}|R><R|\\
{\bar{\cal F}}^{\mu {\bar y}}=&-\big(\p_\mu{\mit\Phi}^\dag+ A_{R}^{\mu}{\mit\Phi}^\dag-{\mit\Phi}^\dag A_{L}^{\mu}\big)|R><L|\\&
-\p^y A_{R}^{\mu}|R><R|-\p^y A_{L}^{\mu}|L><L|\\
{\bar{\cal F}}^{y {\bar y}}=&-\p^y{\mit\Phi}^\dag|R><L|+\p^y{\mit\Phi}|L><R|\\&
+{\mit\Phi}^\dag{\mit\Phi}|R><R|
-{\mit\Phi}{\mit\Phi}^\dag|L><L|
\end{aligned}\label{2-8}
\ee
Then, we define the Lagrangian by use of field strengthes in \eqref{2-6} and \eqref{2-8}.
\begin{align}
&\begin{aligned}
{\cal L}(x,y)=&-\frac{1}{2g_1^2}\text{Tr}<{\bar{\cal F}^{\mu\nu\dag}}(x,y),\; {\cal F}_{\mu\nu}(x,y)>\\&
-\frac{1}{g_2^2}\text{Tr}<{\bar{\cal F}^{\mu y\dag}}(x,y),\; {\cal F}_{\mu y}(x,y)>
\\
&-\frac{1}{g_3^2}\text{Tr}<{\bar{\cal F}^{\mu {\bar y}\dag}}(x,y),\; {\cal F}_{\mu{\bar y}}(x,y)>\\&-\frac{1}{g_4^2}\text{Tr}<{\bar{\cal F}^{y {\bar y}\dag}}(x,y),\; {\cal F}_{y{\bar y}}(x,y)>\\
\end{aligned}\label{2-9}\\
&\hskip1.2cm\begin{aligned}
=&-\frac{1}{2g_1^2}\text{Tr}\,F_{L}^{\mu\nu\dag}F_{L\mu\nu}-\frac{1}{2g_1^2}\text{Tr}\,F_{R}^{\mu\nu\dag}F_{R\mu\nu}\\
&+\left(\frac{1}{g_2^2}+\frac{1}{g_3^2}\right)\text{Tr}\,
\left(\big({\cal D}^\mu{\mit\Phi}\big)^\dag\big({\cal D}_\mu{\mit\Phi}\big)
\right.\\&
-\left.\big(\p^yA_{L}^{\mu}\big)^\dag\big(\p_yA_{L\mu}\big)
-\big(\p^yA_{R}^{\mu}\big)^\dag\big(\p_yA_{R\mu}\big)\right)\\
&+\frac{2}{g_4^2}\text{Tr}\,\left(\big(\p^y{\mit\Phi}\big)^\dag\big(\p_y{\mit\Phi}\big)
-\big({\mit\Phi}^\dag{\mit\Phi}\big)^\dag\big({\mit\Phi}^\dag{\mit\Phi}\big)\right),
\end{aligned}\label{2.6}
\end{align}
where
\be
\begin{aligned}
{F}_L^{\mu\nu}=&\;\p^\mu A_{L}^{\nu}-\p^\nu A_{L}^{\mu}+[A_{L}^{\mu},\,A_{L}^{\nu}],\\
{F}_R^{\mu\nu}=&\;\p^\mu A_{R}^{\nu}-\p^\nu A_{R}^{\mu}+[A_{R}^{\mu},\,A_{R}^{\nu}],\\
{\cal D}^\mu{\mit\Phi}=&\;\p^\mu{\mit\Phi}+ A_{L}^{\mu}{\mit\Phi}-{\mit\Phi} A_{R}^{\mu}.
\end{aligned}
\ee
\par
Here, we address the gauge transformation of the present formulation. The transformation function is denoted by
\be
g(x)=g_L(x)|L><L|+g_R(x)|R><R|,
\ee
in which we should note that $g(x)$ does not depend on the argument $y$ of $S_1$. It is evident that the covariant derivatives are gauge covariant
as they should be.
\be
\begin{aligned}
g(x){\cal D}_\mu g^{-1}(x)=&\:\p_\mu+A^g_{L\mu}|L><L|\\&
+A^g_{R\mu}|R><R|
={\cal D}^g_\mu,\\
g(x){\cal D}_y g^{-1}(x)=&\:
\p_y +{\mit\Phi}^g |L><R|={\cal D}^g_y,\\
g(x){\cal D}_{\bar y}g^{-1}(x)=&\:\p_y+{\mit\Phi}^{g\dag} |R><L|
={\cal D}_{\bar y}^g.
\end{aligned}\label{2-13}
\ee
Similarly, we can prove the covariant derivatives given in \eqref{1.3}
are also covariant for the gauge transformation.
According to \eqref{2-13}, field strengthes expressed in 
\eqref{2-6} and \eqref{2-8} are gauge covariant which yields that
the Lagrangian in \eqref{2-9} is gauge invariant. It is also evident that
the leptonic Lagrangian in \eqref{2-6a} is gauge invariant.
\section{Reconstruction of Weinberg model}
Let us specify  gauge fields in W-S model as 
\be
\begin{aligned}
& A_{L\mu}=-\frac i2\left\{\sum_{k=1}^3\sigma^k g{A_{L}^k}_\mu
                          +a\sigma^0g'B_\mu\right\},\\
& A_{R\mu}=-\frac i2bg'B_\mu,\\[2mm]
&{\mit\Phi}=\begin{pmatrix}
            \phi^+ \\ \phi^0
             \end{pmatrix},
\end{aligned}
\ee
where ${A_{L}^k}_\mu$ and $B_\mu$ are $SU(2)$ and $U(1)$ gauge fields
with coupling constants $g$ and $g'$, 
respectively and, $a$ and $b$ are the $U(1)$ hypercharges of left- 
and right-handed leptons, respectively. From \eqref{2-6}, we can form the
fields strengthes
\be
\begin{aligned}
{F}_L^{\mu\nu}=&-\frac i2g\sum_{i=1}^3\sigma^i \big(\p^\mu A_L^{i\nu}
-\p^\nu A_L^{i\mu}+gf^{ijk}A_L^{j\mu}A_L^{k\nu}\big)\\&
-\frac i2ag'\sigma^0
\big(\p^\mu B^{\nu}
-\p^\nu B^{\mu}\big),\\
{F}_R^{\mu\nu}=&-\frac i2bg'\sigma^0
\big(\p^\mu B^{\nu}-\p^\nu B^{\mu}\big),\\
{\cal D}^\mu{\mit\Phi}=&\left\{\p^\mu
-\frac i2\left(\sum_{k=1}^3\sigma^k g{A_{L\mu}^k}
                          -\sigma^0g'B_\mu\right)\right\}{\mit\Phi}.
\end{aligned}\label{2.9}
\ee
After insertion of \eqref{2.9} into \eqref{2.6} and rescaling of fields,
the Lagrangian takes the form, with constant parameters $\alpha,\,\beta_L,\,\beta_R$ and $\lambda'$ resulting from proper calculation
\be
\begin{aligned}
{\cal L}(x,y)=&-\frac14\sum_{i=1}^3F_L^{i\mu\nu}F_{L\mu\nu}^i
-\frac14B^{\mu\nu}B_{\mu\nu}\\
+\big({\cal D}^\mu{\mit\Phi}\big)^\dag&\big({\cal D}_\mu{\mit\Phi}\big)
-\lambda'\big({\mit\Phi}^\dag{\mit\Phi}\big)^2
+\alpha^2\big(\p^y{\mit\Phi}\big)^\dag\big(\p_y{\mit\Phi}\big)\\
-\beta_L^2\sum_{i=1}^3\big(&\p^yA_L^{i\mu}\big)\big(\p_yA_{L\mu}^{i}\big)
-\beta_R^2\big(\p^yB^{\mu}\big)\big(\p_yB_{\mu}\big),
\end{aligned}
\ee
where
\be
\begin{aligned}
& F^i_{L\mu\nu}=\p_\mu{A_{L}^i}_\nu-\p_\nu{A_{L}^i}_\mu +gf_{ijk}{A_{L}^j}_\mu{A_{L}^k}_\nu,\\
& B_{\mu\nu}=\p_\mu B_\nu-\p_\nu B_\mu,\\[0mm]
&{\cal D}_{\mu}{\mit\Phi}=\left\{\p_\mu -\frac{i}2\left(\sum_{k=1}^3\sigma^k g{A_{L}^k}_\mu+\sigma^0g'B_\mu
\right)\right\}{\mit\Phi},
\end{aligned}
\ee
with $a=-1$ and $b=-2$ for left- and right-handed leptons in \eqref{2-2a}, respectively.
\par
In this paper, we consider the case that only Higgs field is infiltrated into 
the extra dimensional space such as
\be
{\mit\Phi}(x,y)=\phi(x)f(y),
\ee
where $f'(y)\ne0$ and we normalize the function $f(y)$ such as 
\be
 \frac{1}{2\pi R}\int_0^{2\pi R}f^2(y)dy=1,
\ee
whereas gauge fields  penetrate the extra dimension uniformly and therefore don't depend on the argument y of extra dimensional space.
\be
A_{L}^i(x,y)=A_{L}^i(x),\hskip0.5cm\,B_\mu(x,y)=B_\mu(x).
\ee
From these settings, we find the 4-dimensional Yang-Mills-Higgs Lagrangian
\begin{align}
L_{Y\!M\!H}&=\frac{1}{2\pi R}\int_0^{2\pi R}{\cal L}(x,y)dy\nonumber\\
&\hskip-8mm
\begin{aligned}
=&-\frac1{4}\sum_{i=1}^3F^i_{L\mu\nu}(x)F^{i\mu\nu}_{L}(x)
-\frac14B_{\mu\nu}(x)B^{\mu\nu}(x)\\
&+({\cal D}_\mu{\phi}(x))^\dag({\cal D}^\mu{\phi}(x))\\&
+\frac{\alpha^2}{2\pi R}\int_0^{2\pi R}{f'}^{2}(y)dy\,\big({\phi}^\dag(x){\phi}(x)\big)\\
&-\frac{\lambda'}{2\pi R}\int_0^{2\pi R}f^4(y)dy\,\left({\phi^\dag}(x){\phi}(x)\right)^2,
\end{aligned}
\label{3.8}
\end{align}
where we should notice that  the metric structure is
\be
\p_y=\p^y.
\ee
The effective potential in tree level is known  to be
\be
V(\phi)=\lambda\left({\phi^\dag}(x){\phi}(x)\right)^2-\mu^2\big({\phi}^\dag(x){\phi}(x)\big)
\ee
with the parameters $\lambda=\frac{\lambda'}{2\pi R}\int_0^{2\pi R}f^4(y)dy$ 
and $\mu^2=\frac{\alpha^2}{2\pi R}\int_0^{2\pi R}{f'}^{2}(y)dy$.
Here, we adopt the unitary gauge and then Higgs field is expressed as
\begin{align}
&{\phi}=\begin{pmatrix}
0 \\ \dfrac{\varphi+v}{\sqrt{2}}
\end{pmatrix},
\end{align}
where 
${\phi}^0=\left(
0 \quad {v}/{\sqrt{2}}
\right)^t
$ 
gives the minimal point to the effective potential $V(\phi)$, and so 
$v=\mu/\sqrt{\lambda}$. The field $\varphi$ is the neutral Higgs boson. The effective potential as a function of $\varphi$ 
is
\be
V(\varphi)=\frac{\lambda}{4}\varphi^4+\lambda v\varphi^3+\mu^2\varphi^2
\ee
except for the constant term. The covariant derivative of $\phi$ is written as
\begin{align}
{\cal D}_\mu\phi=
&\frac{1}{\sqrt{2}}
\begin{pmatrix} 0 \\ \p_\mu\varphi \end{pmatrix}
-\frac{i}{2}\begin{pmatrix}
 \sqrt{2}gW_\mu^+ \\[3mm]
 \sqrt{g^2+g^{'2}}Z_\mu
\end{pmatrix}\frac{\varphi+v}{\sqrt{2}},
\end{align}
where $W_\mu^+$ and $Z_\mu$ are the charged and neutral weak boson fields, respectively. Finally, we obtain the Yang-Mills-Higgs Lagrangian
\be
\begin{aligned}
L_{Y\!M\!H}=&-\frac1{4}\sum_{i=1}^2F^i_{L\mu\nu}(x)F^{i\mu\nu}_{L}(x)
\\&
-\frac14F_{Z\mu\nu}(x)F_Z^{\mu\nu}(x)-\frac14F_{\mu\nu}(x)F^{\mu\nu}(x)\\
&+\frac12(\p_\mu\varphi)^2-V(\varphi)\\
+\frac14(\varphi&+v)^2\left(g^2W_\mu^+ W^{-\mu}+
\frac{(g^2+g^{'2})}{2}Z^2_\mu\right)
,
\end{aligned}
\label{3.13}
\ee
where $F^i_{L\mu\nu},\;F_Z^{\mu\nu}$ and $F_{\mu\nu}(x)$ are the field strengths of charged, neutral weak gauge fields and photon field, respectively.
From \eqref{3.13}, the famous mass relation $m_W=m_Z\cos \theta_W$ follows.\par
Under the assumption that  leptons also stay at $S_1$ uniformly, 
the 4-dimensional Dirac Lagrangian obtained by integrating \eqref{2-6a} 
takes the form
\be
\begin{aligned}
L_D=&\frac{1}{2\pi R}\int_0^{2\pi R}{\cal L}_D\,dy\\
={\bar{\psi}}_L&i\gamma^\mu\left\{\p_\mu-\frac i2\left(\sum_{k=1}^3\sigma^k g{A_{L}^k}_\mu-\sigma^0g'B_\mu\right)\right\}\psi_L \\
+{\bar{e}_R}&i\gamma_\mu\left(\p^\mu+ ig'B_\mu\right)e_R
+g_Y{\bar e} e\varphi+m_e{\bar e}e,
\end{aligned}
\label{3.15}
\ee
where 
$g_Y=\frac{g'_Y\sqrt{\lambda'}}{2\sqrt{2}\alpha\pi R}\int_0^{2\pi R}f(y)dy$
 and the electron mass $m_e=g_Yv$.
The equation \eqref{3.15} is equal to the lepton part of  Lagrangian in the Weinberg-Salam model.
\section{Conclusions}
We have reconstructed the Weinberg-Salam model based on the generalized covariant derivative method on the product space $M_4\times S_1$, where the gauge symmetry is spontaneously broken owing to the penetration of the Higgs field into the extra compact space $S_1$. This breakdown of symmetry is realized without considering the quantum effects. This is favorable point of our model. 
\par
It is assumed that the gauge  and fermion fields do not depend on the argument  $y$ in $S_1$, and therefore the Kaluza-Klein modes of those fields do not appear on the stage. This assures the renormalizability and stability of our model formulated in this paper since the $y$ derivative terms of gauge fields in \eqref{2.6}
give the imaginary masses to the Kaluza-Klein modes.
The existence of the Kaluza-Klein modes of Higgs field is also undesirable
since it increases the gauge boson mass.
This non-existence of  Kaluza-Klein modes is a clear difference from other model \cite{Hosotani}.
LHC or more powerful machine 
will decide whether Kaluza-Klein modes exists or not.
\bigskip

\bigskip
\begin{center}
{\bf Acknowledgement}
\end{center}
\smallskip
The author would like to
express his sincere thanks to
Professor
 H.~Kase and Professor K. Morita 
for useful suggestion and
invaluable discussions.

\end{document}